\documentclass[a4paper,11pt]{article}
\usepackage{pos}

\usepackage{xcolor}
\usepackage{amsmath}
\usepackage{slashed}
\usepackage{stmaryrd}%\lightning
\usepackage{bbm}
\usepackage{braket}
\usepackage{wrapfig}

\newcommand{\mui}{\mu_I}
\newcommand{\refc}[1]{(\ref{#1})}

\title{QCD thermodynamics at non-zero isospin asymmetry}%\thanks{}

\ShortTitle{QCD thermodynamics at non-zero isospin asymmetry}

\author*[a]{Bastian B. Brandt}
\author[b]{Francesca Cuteri}
\author[a]{Gergely Endr\H{o}di}

\affiliation[a]{Institute for Theoretical Physics, University of Bielefeld, D-33615 Bielefeld, Germany}

\affiliation[b]{Institute for Theoretical Physics, Goethe University, D-60438 Frankfurt am  Main, Germany}

\emailAdd{brandt@physik.uni-bielefeld.de}
\emailAdd{endrodi@physik.uni-bielefeld.de}
\emailAdd{cuteri@itp.uni-frankfurt.de}

\abstract{We study the thermodynamic properties of QCD at nonzero isospin chemical potential using improved staggered quarks at physical quark masses. In particular, we discuss the determination of the equation of state at zero and nonzero temperatures and show results. Using the results for the isospin density $n_I$, we also determine the phase diagram in the $(n_I,T)$-plane.}

\FullConference{%
 The 38th International Symposium on Lattice Field Theory, LATTICE2021
  26th-30th July, 2021
  Zoom/Gather@Massachusetts Institute of Technology
}

\begin{document}
\maketitle

\section{Introduction}

The QCD phase diagram and the equation of state (EoS) at non-zero quark densities have been the subject of intense experimental and theoretical research in the past two decades. Physical systems typically feature a non-zero charge and strangeness chemical potential on top of the baryon chemical potential. The effect of the latter is usually dominant, but the other chemical potentials can potentially also contribute a sizeable and important contribution to the dynamics of the system. Furthermore, in some cases the charge chemical potential can play the major role. This is the case for an early Universe starting with large lepton flavour asymmetries~\cite{Wygas:2018otj,Middeldorf-Wygas:2020glx,Vovchenko:2020crk}. Non-zero charge chemical potential directly translates into a non-zero isospin chemical potential $\mu_I$. While lattice simulations with a generic combination of quark chemical potentials suffer from the infamous complex action (or sign) problem, QCD at pure isospin chemical potential, i.e. at vanishing other chemical potential components, has a real action and is amenable to direct Monte-Carlo simulations.

\begin{wrapfigure}{r}{7.2cm}
 \centering
 \vspace*{-3mm}
 \includegraphics[width=7cm]{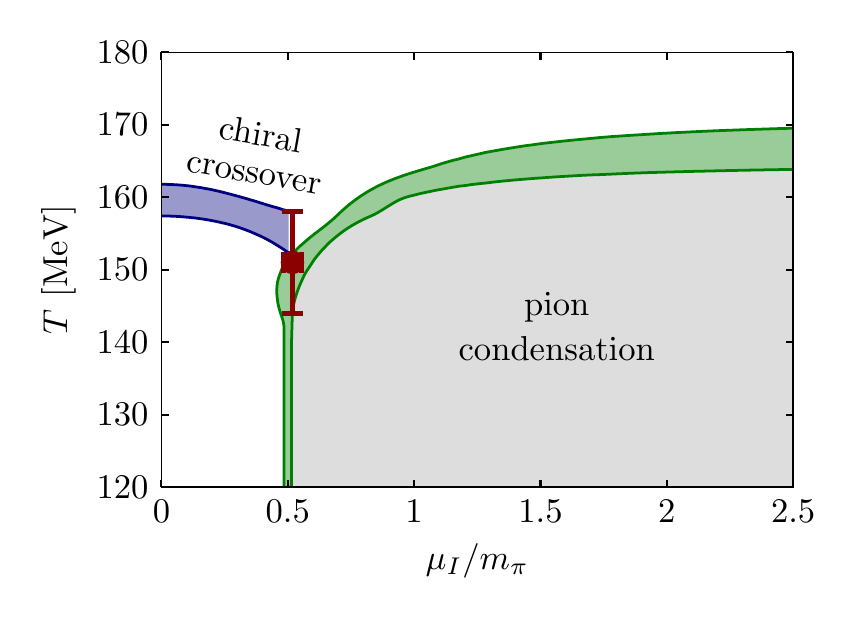}
 \caption{Continuum phase diagram of QCD at pure isospin
 chemical potential~\cite{Brandt:2017oyy,Brandt:2018omg} (taken from Ref.~\cite{Brandt:2018wkp}).}
 \label{fig:phdiag}
\end{wrapfigure}

In the past few years we have performed extensive studies of QCD at non-zero isospin asymmetry. In particular, we have computed the continuum phase diagram at physical pion mass~\cite{Brandt:2017oyy,Brandt:2018omg}, first discussed in chiral perturbation theory in Ref.~\cite{Son:2000xc}~\footnote{First pioneering studies on the phase structure have been reported in Refs.~\cite{Kogut:2002tm,Kogut:2002zg,Kogut:2004zg,deForcrand:2007uz,Detmold:2012wc,Cea:2012ev}.}, featuring the standard hadronic and quark-gluon plasma (QGP) phases and a phase with a Bose-Einstein condensate of charged pions (BEC phase). The phase diagram is shown in Fig.~\ref{fig:phdiag}. Furthermore, perturbation theory predicts the existence of a superconducting BCS phase at asymptotically large $\mu_I$~\cite{Son:2000xc} and we are currently investigating whether this phase could already be present for the intermediate chemical potentials at our disposal, see Ref.~\cite{Brandt:2019hel} and the contribution to these proceedings Ref.~\cite{Brandt:DiracLat2021}.

In this proceedings article we discuss the extraction of the EoS at non-zero $\mu_I$ and present results for different temperatures and $N_t=8,\,10$ and 12 lattices. First accounts on our study of the EoS have been given in Refs.~\cite{Vovchenko:2020crk,Brandt:2017zck,Brandt:2018wkp}. Furthermore, we present the phase diagram in the plane of isospin density and temperature, $(n_I,T)$-plane, for which we perform a tentative continuum extrapolation.

\section{The equation of state at non-zero isospin asymmetry}

The extraction of the EoS at $T=0$ has so far been done on a single lattice spacing only. The details have already been published in Ref.~\cite{Brandt:2018bwq}. To extract the EoS at $\mu_I\neq0$ and $T\neq0$, we use the ensembles of Ref.~\cite{Brandt:2017oyy}, which have already been used to map out the phase diagram up to $\mu_I/m_\pi\lesssim 0.9$. We refer the interested reader to this reference for the simulation details. The ensembles we use entail three different temporal extents, $N_t=8,\,10$ and 12, with an aspect ratio $N_s/N_t\approx 3$, and are set up with physical mass $u$, $d$ and $s$ quarks ($u$ and $d$ are mass-degenerate). To simulate this setup, we use improved rooted staggered fermions with two levels of stout smearing. Simulations are done at non-vanishing regulator $\lambda$ (pionic source) and results are extrapolated to $\lambda=0$ using the improvement program described in Ref.~\cite{Brandt:2017oyy}. The most important observable for this proceedings article is the isospin density $n_I$, for which the definition and extrapolation details are discussed in Ref.~\cite{Brandt:2018bwq}. From now on we will only use quantities which have already been extrapolated to $\lambda=0$.

\subsection{Extracting the EoS from the isospin density}

To compute the EoS from the lattice, the main task is the computation of the isospin density, pressure $p$ and the interaction measure $I$, which determine the other related quantities. For energy and entropy density, $\epsilon$ and $s$, for instance, we have
\begin{equation}
\label{eq:ene-ent}
 \epsilon = I+3p \quad \text{and} \quad s=\frac{\epsilon + p - \mu_I n_I}{T} \,.
\end{equation}
The isospin density can be computed directly from the simulations, see Ref.~\cite{Brandt:2018bwq}. What remains is the computation of $p$ and $I$. It is convenient to rewrite these quantities as
\begin{equation}
 p(T,\mu_I) = p(T,0) + \Delta p(T,\mu_I) \quad \text{and} \quad I(T,\mu_I) = I(T,0) + \Delta I(T,\mu_I) \,.
\end{equation}
The $\mu_I=0$ contributions are already available in the literature, e.g.~\cite{Borsanyi:2013bia,HotQCD:2014kol}. For our action results are also available in Ref.~\cite{Borsanyi:2010cj}. This leaves the computation of the modifications of the EoS due to the isospin chemical potential, $\Delta p$ and $\Delta I$. The computation of the modification of the pressure has already been discussed in Refs.~\cite{Vovchenko:2020crk,Brandt:2018wkp,Brandt:2017zck}. It can be computed using
\begin{equation}
\label{eq:plat}
 \Delta p(T,\mu_I) = \int_0^{\mu_I} d\mu \, n_I(T,\mu) \,.
\end{equation}
The computation of the interaction measure uses the relation
\begin{equation}
\label{eq:intmeas}
 \frac{I(T,\mu)}{T^4} = T \frac{\partial}{\partial T} \left(\frac{p(T,\mu_I)}{T^4}\right)
+ \frac{\mu_I n_I(T,\mu_I)}{T^4} \,.
\end{equation}
There are two basic possibilities to compute $I$: (a) 
rewrite it as a derivative of the partition function with respect to the lattice scale
%use the extension of the integral method~\cite{Engels:1990vr} (see also Ref.~\cite{Borsanyi:2010cj}) 
at non-zero chemical potentials (see, e.g., Ref.~\cite{Allton:2003vx}) and evaluate the relevant observables using $\mu_I=0$ subtraction; (b) use a two-dimensional interpolation of the results for $n_I$ to obtain the function $n_I(T,\mu_I)$ and use Eq.~\eqref{eq:intmeas}. Both have different systematics and challenges and eventually one wants to compare the results from the two methods. Overall we found a careful implementation of method (b) to lead to more accurate results, which we will present in the following.

To be able to evaluate $I$, we can insert the pressure from Eq.~\refc{eq:plat}, which results in (see also Ref.~\cite{Vovchenko:2020crk})
\begin{equation}
 \Delta I(T,\mui) = \mui n_I(T,\mui) + \int_0^{\mui} d\mui' \Big[ T \frac{\partial}{\partial T} - 4 \Big] n_I(T,\mui') \,.
\end{equation}
Thus, given a suitable differentiable and integrable interpolation for $n_I(T,\mu_I)$, we can compute the interaction measure using this expression.

\subsection{Interpolation of the isospin density}

The remaining task is performing the interpolation, ideally in a way which is as model-independent as possible. Note, that an interpolation of an unknown function based on a discrete set of points with uncertainties is an ill-posed inverse problem and, as such, does not have a unique solution. So one of the tasks is to find a solution which is close to the actual physical solution. This is already true for the interpolations of the EoS at $\mu_I=0$ and will remain true through the necessary interpolation for $n_I$ at $\mu_I\neq0$. Here we will try to remain as model-independent as possible by using all possible two-dimensional spline interpolations which provide a ``good'' description of the data. These spline interpolations are generated by a spline Monte-Carlo discussed in Ref.~\cite{Brandt:2016zdy}. The included weight function, as well as the imposed boundary conditions are chosen carefully to reduce model dependence as much as possible. In addition, the spline boundary conditions can be used to include additional physical information on the function at hand.

\begin{wrapfigure}{r}{8cm}
 \centering
 \vspace*{-3mm}
 \includegraphics{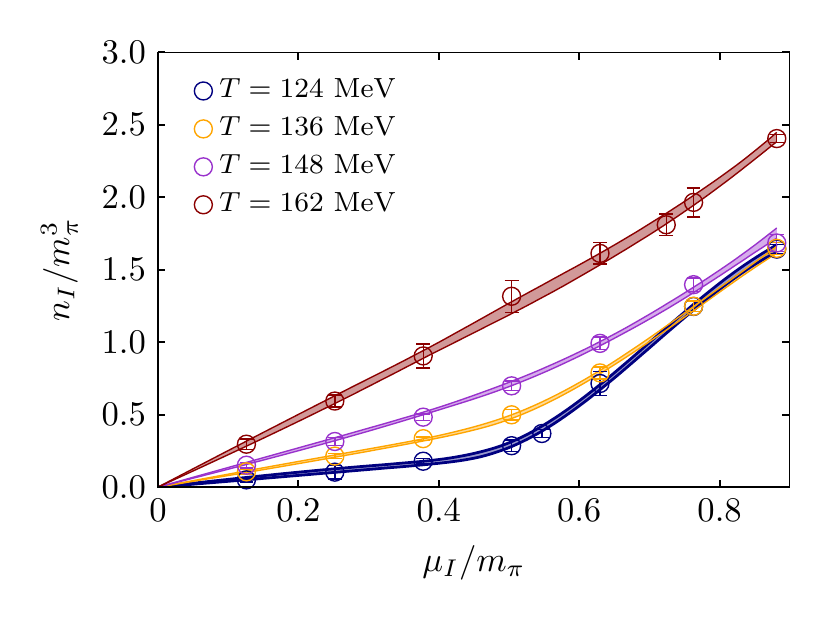}
 \caption{Two-dimensional interpolation of the isospin density on the $N_t=8$ ensembles as explained in the text.}
 \label{fig:niinterp}
\end{wrapfigure}

For the interpolation of $n_I$, we use two-dimensional cubic splines for which we set the outer spline grid point at small $\mu_I$ to $\mu_I=0$ and impose the physical conditions that $n_I$ and its second derivative are zero on the $\mu_I=0$ axis ($n_I$ is an odd function of $\mu_I$). The position of the other outer grid point in $\mu_I$-direction and the outer grid points in $T$-direction is variable and we keep the second derivatives at the outer points as free parameters (note, that all other boundary conditions are determined by the interpolation). In the weight for the Monte-Carlo average we use the Akaike information criterion as action (see Ref.~\cite{Brandt:2016zdy} and references therein for details) and include another term which is designed to suppress oscillatory solutions (see section 4 in Ref.~\cite{Endrodi:2010ai}).\footnote{In practice, we have used $\mathcal{D}$ as defined in eq. (15) in Ref.~\cite{Endrodi:2010ai}, replacing $f$ in the denominators by its statistical uncertainty and with $\epsilon$ set according to a small fraction of the width between the datapoints closest to the varied grid point.} The associated free parameter for a relative weighting of the two terms is kept as small as possible, while not leading to solutions which oscillate strongly.

The resulting interpolation on a $N_t=8$ lattice is shown for a set of temperatures in Fig.~\ref{fig:niinterp}. The uncertainties include the uncertainty due to the individual data points (computed using the bootstrap procedure with 1000 samples) and from the Monte-Carlo over spline interpolations.

\subsection{Results for the EoS}

\begin{figure}[t]
 \centering
 \begin{minipage}{.48\textwidth}
 \centering
 \includegraphics[width=\textwidth]{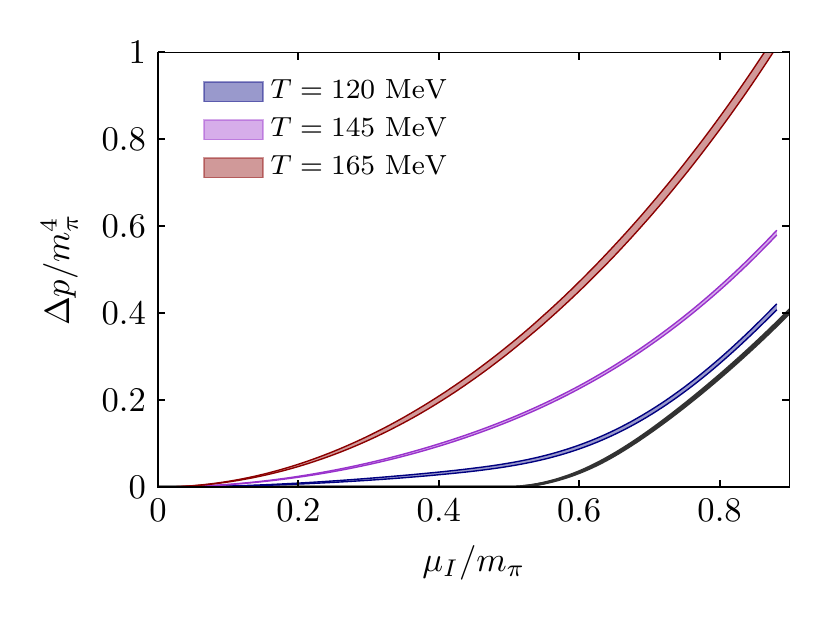} \\
 \includegraphics[width=\textwidth]{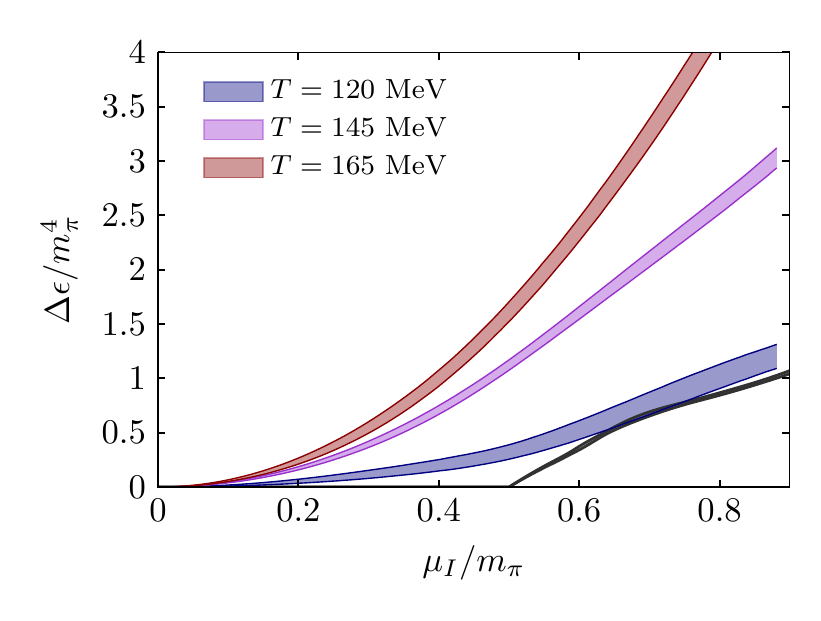}
 \end{minipage}
 \begin{minipage}{.48\textwidth}
 \centering
 \includegraphics[width=\textwidth]{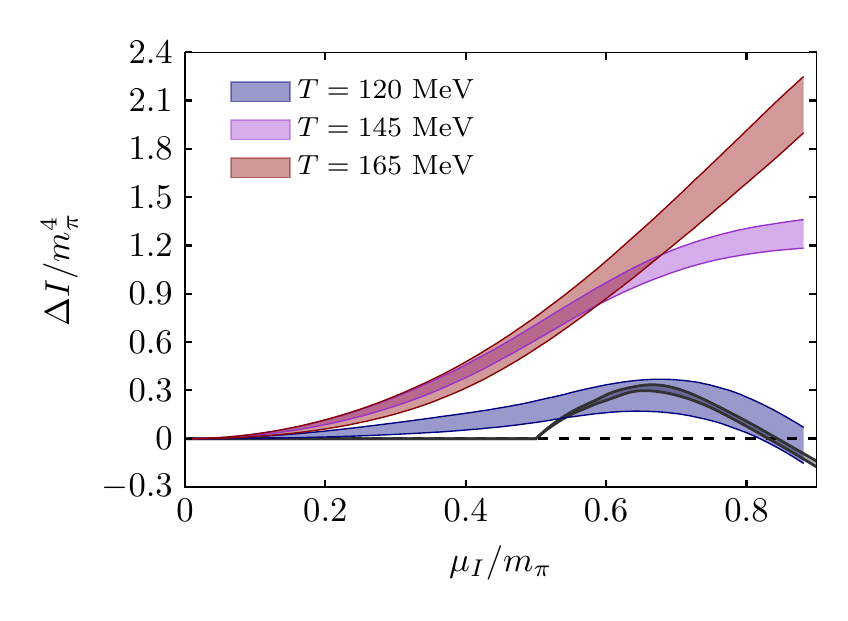} \\
 \includegraphics[width=\textwidth]{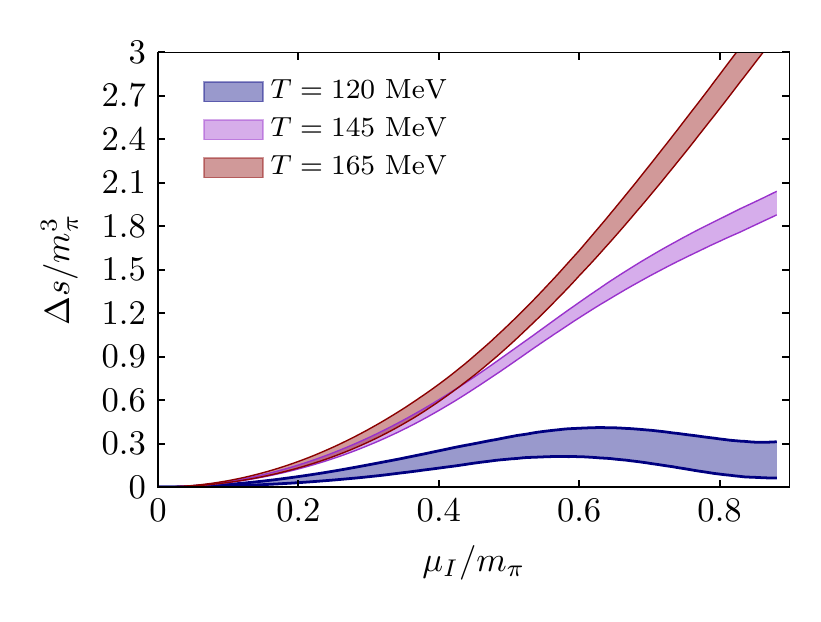}
 \end{minipage}
 \caption{
 Results for the pressure (top-left), the interaction measure (top-right), the energy density (bottom-left) and and the entropy density (bottom-right). The results have been obtained on our lattice with $N_t=8$ from the spline interpolation of the isospin density discussed in the main text. The dark gray curves are the $T=0$ results from Ref.~\cite{Brandt:2018bwq}.}
 \label{fig:eos-nt8}
\end{figure}

Given the spline interpolation discussed in the previous section, we can now compute $\Delta p$ and $\Delta I$ from Eqs.~\refc{eq:plat} and~\refc{eq:intmeas}, evaluating the derivatives and integrals analytically. Using Eq.~\refc{eq:ene-ent} we further evaluate the energy and entropy densities. The results are shown for different temperatures and the $N_t=8$ ensemble in Fig.~\ref{fig:eos-nt8}. We also show the $T=0$ results from Ref.~\cite{Brandt:2018bwq} (dark gray curves). Evidence for the presence of the pion condensate in the EoS is given by the characteristic behavior of the interaction measure (top-right). At $T=0$ it initially increases starting at the BEC phase boundary, $\mu_I/m_\pi=0.5$, before reaching a maximum at around $\mu_I/m_\pi\approx 0.66$ from where it decreases and becomes negative at around $\mu_I/m_\pi\approx 0.84$. Note, that for $T=0$ the interaction measure and the other quantities vanish at $\mu_I=0$. These findings are in agreement with the results from chiral perturbation at next-to-leading order~\cite{Adhikari:2019zaj}. The same behavior can still be seen for $T=120$~MeV and to some extent for $T=145$~MeV, where, however, the maximum (if it still exists) is shifted towards larger values of $\mu_I$. For $T=165$~MeV the system hardly enters the BEC phase and no sign of pion condensation is present in $\Delta I$. The BEC phase also leads to non-monotonous behavior of the entropy density with $\mu_I$ at low temperatures.

\begin{figure}[t]
 \centering
 \begin{minipage}{.48\textwidth}
 \centering
 \includegraphics[width=\textwidth]{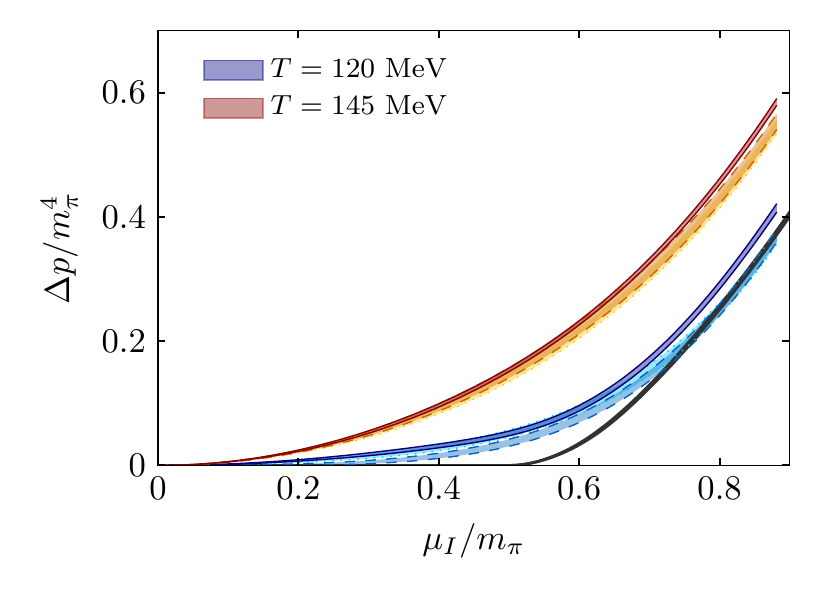}
 \end{minipage}
 \begin{minipage}{.48\textwidth}
 \centering
 \includegraphics[width=\textwidth]{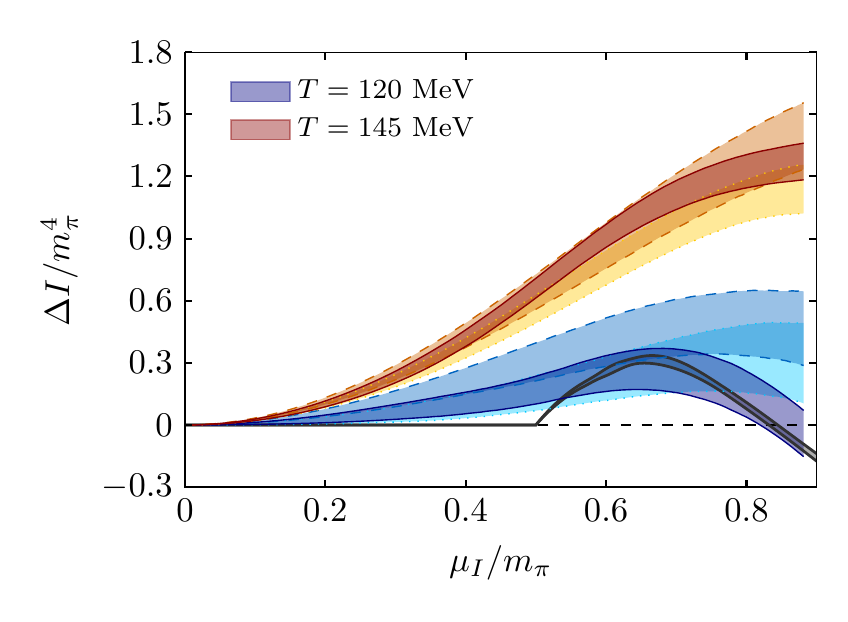}
 \end{minipage}
 \caption{Comparison of the results for pressure (left) and interaction measure (right) for different values of $N_t$. The bands with the solid bounding lines are the ones for $N_t=8$ as shown already in Fig.~\ref{fig:eos-nt8}, the ones with the shaded bounding lines the ones for $N_t=10$ and the ones with dotted bounding lines the ones for $N_t=12$.}
 \label{fig:I3nt}
\end{figure}

Similar results are available for the $N_t=10$ and 12 lattices and we show a comparison between the results for pressure and interaction measure in Fig.~\ref{fig:I3nt}. While for the interaction measure the bands overlap in most regions of parameter space, the pressure shows clear signs for sizeable lattice artifacts at $N_t=8$. This is not unexpected given the large lattice artifacts for the leading order Taylor coefficient at $N_t=8$ observed in Ref.~\cite{Borsanyi:2011sw} together with the fact that the Taylor expansion provides a good description for the data at small $\mu_I$~\cite{Brandt:2018omg} and the accumulation of differences in $n_I$ in the integration to obtain $\Delta p$, see Eq.~\refc{eq:plat}. Consequently, $N_t=8$ is not yet within the scaling region for a leading order continuum limit, for which $N_t=16$ lattices are needed for a reliable extrapolation.

\section{The phase diagram in the {\boldmath $(n_I,T)$}-plane}

\begin{figure}[b]
 \centering
 \includegraphics[width=.5\textwidth]{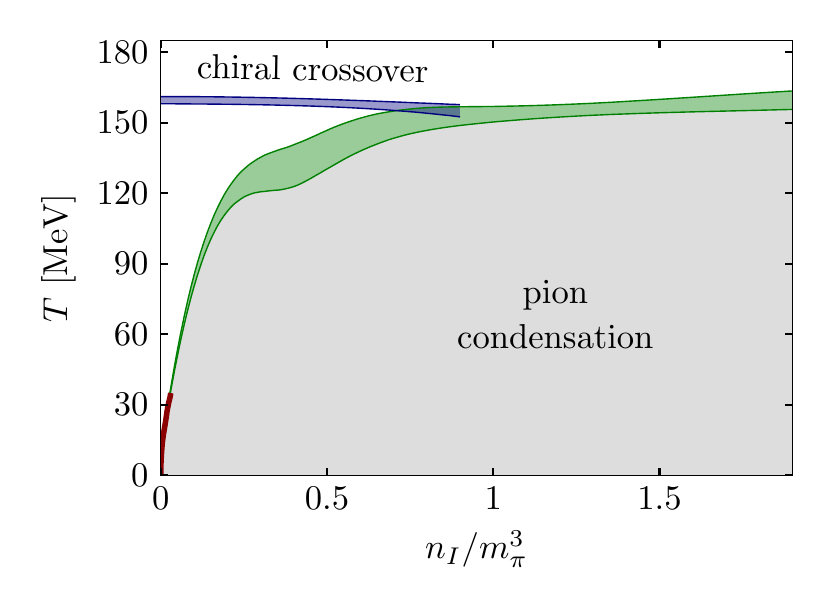}
 \caption{Tentative continuum extrapolation for the phase diagram in the $(n_I,T)$-plane, as explained in the main text. The red line corresponds to the next-to-leading chiral perturbation theory prediction at small temperatures~\cite{Adhikari:2020kdn}.}
 \label{fig:niTphd}
\end{figure}

With the interpolation for $n_I$ we can also determine the phase diagram in the $(n_I,T)$-plane. A tentative continuum extrapolation of the phase diagram using the $N_t=10$ and 12 results only (for the same reasoning as above we do not expect $N_t=8$ to be in the scaling region) is shown in Fig.~\ref{fig:niTphd}. Note, that lattice results are not available below $T=120$~MeV, but we have matched the continuum extrapolation to next-to-leading order chiral perturbation theory at a temperature of 30~MeV~\cite{Adhikari:2020kdn}, shown in Fig.~\ref{fig:niTphd} as the red line. We note, that the $N_t=10$ and 12 phase boundaries are fully included in the uncertainties for the continuum phase boundaries. A final continuum extrapolation will be obtained once the $N_t=16$ results become available.

\section{Conclusions}

We have shown first results for the full equation of state at non-zero isospin chemical potential from $N_t=8,\,10$ and 12 ensembles. The results have been obtained from a two-dimensional spline interpolation of the isospin density with reduced model dependence due to a spline Monte-Carlo analysis. The interaction measure shows a characteristic behavior with $\mu_I$ due to the presence of pion condensation with an initial rise before reaching a miximum and decreasing. For small temperatures it eventually becomes negative. This behavior is also seen in chiral perturbation at next-to-leading order~\cite{Adhikari:2019zaj}. Using the interpolation for the isospin density, we have also determined the phase diagram in the $(n_I,T)$-plane and performed a tentative continuum extrapolation matching to next-to-leading order chiral perturbation theory at small temperatures. For the isospin density, the $N_t=8$ results seem to lie outside of the scaling region. Consequently results for $N_t=16$ are required for reliable continuum extrapolations and we are currently extending our set of ensembles accordingly. \\

\noindent\textbf{Acknowledgements:}\\
We thank Prabal Adhikari, Jens Oluf Andersen and Martin Mojahed for discussions and for providing the chiral perturbation theory data from Ref.~\cite{Adhikari:2020kdn}. This work has been supported by the Deutsche Forschungsgemeinschaft (DFG, German Research Foundation) via the Emmy Noether Programme EN 1064/2-1 and TRR 211 – project number 315477589. The authors gratefully acknowledge the Gauss Centre for Supercomputing e.V. (\href{https://www.gauss-centre.eu}{\tt www.gauss-centre.eu}) for funding this project by providing computing time on the GCS Supercomputer SuperMUC-NG at Leibniz Supercomputing Centre (\href{https://www.lrz.de}{\tt www.lrz.de}).

\providecommand{\href}[2]{#2}\begingroup\raggedright\endgroup

% \bibliography{lat21_references}
% \bibliographystyle{JHEP}

\end{document}